\documentclass[prl,twocolumn,amsmath,amssymb,aps]{revtex4-2}
\usepackage{graphicx,subfigure,amsmath,bm,colordvi,times}
\usepackage{mathtools}
\usepackage[usenames]{color}
\usepackage{multirow}
\usepackage{epstopdf}
\usepackage[utf8]{inputenc}
\usepackage[english]{babel}
\usepackage{amsfonts}
\usepackage{amssymb}
\usepackage{float}
\usepackage{hyperref}
\usepackage{braket}

\hypersetup{pdfborder=0 0 0,colorlinks=true,citecolor=blue,linkcolor=blue}

\begin{document}

\title{
Chester supersolid of spatially indirect excitons in double-layer semiconductor heterostructures
}
\author{Sara Conti$^{1}$, Andrea Perali$^2$, Alexander R. Hamilton$^3$,
 Milorad V. Milo\u sevi\'c$^{1,4}$, Fran\c{c}ois M. Peeters$^{1}$, and David Neilson$^{1,3}$}
\affiliation{
$^1$Department of Physics, University of Antwerp, 2020 Antwerp, Belgium\\
$^2$Supernano Laboratory, School of Pharmacy, University of Camerino, 62032 Camerino (MC), Italy\\
$^3$ARC Centre of Excellence for Future Low Energy Electronics Technologies, School of Physics, University of New South Wales, Sydney 2052, Australia\\
$^4$NANOlab Center of Excellence, University of Antwerp, 2020 Antwerp, Belgium
}

\begin{abstract}
A supersolid, a counter-intuitive quantum state in which a rigid lattice of particles flows without resistance, has to date not been unambiguously realised. 
Here we reveal a supersolid ground state of excitons in a double-layer semiconductor heterostructure over a wide range of layer separations outside the focus of recent experiments.
This supersolid conforms to the original Chester supersolid with one exciton per supersolid site, as distinct from the alternative version reported in cold-atom systems of a periodic modulation of the superfluid density.
We provide the phase diagram augmented by the supersolid. 
This new phase appears at layer separations much smaller than the predicted exciton normal solid, and it persists up to a solid--solid transition where the quantum phase coherence collapses.
The ranges of layer separations and exciton densities in our phase diagram are well within reach of the current experimental capabilities.
\end{abstract}

\maketitle

The existence of supersolid phases has attracted interest for a long time, as it is intriguing and rather counter-intuitive to attempt to visualize particles flowing without resistance while they form a rigid lattice \cite{Leggett1970}.
In this exotic phase, spatial off-diagonal long-range order and periodic solid order coexist, spontaneously breaking particle conservation and continuous translational invariance
\cite{Balibar2008}.
Chester \cite{Chester1970} originally proposed a supersolid ground state of $^4$He, with a single atom localized at each lattice site of the $^4$He crystal. 
There has been some evidence for this phase in $^4$He in torsional-oscillator experiments from non-classical rotational inertia \cite{Kim2004}, but it appears that condensate fractions will be disappointingly tiny \cite{Galli2005}. 
 
An alternative approach to forming a supersolid has involved quantum gases of cold atoms in optical lattices \cite{Ancilotto2019}, and this has renewed interest following recent reports of observations of supersolid phases \cite{Tanzi2019,Guo2019, Natale2019,Bottcher2019}.
However it is important to make the distinction between these periodic density-modulated condensates in cold atoms, with the Chester concept of supersolidity in which single particles quantum condense, while simultaneously they are localized on their individual lattice sites by strong interparticle repulsion. 

Here we show that excitons in a semiconductor heterostructure can form a supersolid of the Chester type, with a single exciton at each site of the supersolid lattice. 
The heterostructure consists of parallel $p$-doped and $n$-doped conducting layers. 
The electrons and holes are spatially confined in their layers by an insulating barrier of thickness $d$ and dielectric constant $\epsilon$. 
The equal carrier densities $\rho$ can be tuned by top and bottom metal gates.
When the average separation between carriers in each layer is much larger than $d$, the electrons and holes will form bound excitonic-like states aligned perpendicular to the layers.
Unlike cold atoms, solidification in this system is driven purely by the repulsion between excitons, the strength of which is tunable by $d$, $\rho$, and $\epsilon$.
We seek supersolidity at low densities and large layer separations where the exciton-exciton repulsion is strong.

This requirement is realizable in a variety of existing semiconductor systems that are the subject of intense experimental interest due to the accumulating evidence that they support electron-hole superfluidity and Bose-Einstein Condensation, 
such as double monolayer transition metal dichalcogenides (TMD) \cite{Conti2020b,Wang2019}, Si/Ge heterojunctions \cite{Conti2021}, double bilayer graphene \cite{Perali2013,Burg2018}, and double quantum wells in III-V semiconductor heterostructures \cite{Stern2014, SaberiPouya2020}. 
Much effort in these systems has focused on achieving very small layer separations where the exciton binding energy is large and the exciton-exciton interaction is weak. 
In contrast, our supersolid is formed at larger interlayer spacing, which is much more experimentally accessible.

We determine the zero-temperature phase diagram as a function of the system parameters, the layer separation $d$ and the density $\rho$ with characteristic length $r_0=1/\sqrt{\pi \rho}$\,. 
The Hamiltonian for the electron-hole pairs is 
\begin{equation} 
H = \sum_{i=1}^N \left(-\frac{\hbar^2}{2M_{\text{X}}}\right)
\mathbf{\nabla}_i^2 +
\sum_{i<j=1}^N V_{\text{XX}}(|\mathbf{r}_i-\mathbf{r}_j|) \, .
\label{Eq:Ham}
\end{equation} 
The index $i$ labels the $N$ exciton pairs of mass $M_{\text{X}}$ at positions $\mathbf{r}_i$ parallel to the layers.
We take equal electron and hole masses, $m_e^*=m_h^*$.
The effective interaction $V_{\text{XX}}(r)$ between two exciton pairs contains the Coulomb interactions acting between the electrons and holes forming the exciton pairs.
We take for the effective exciton-exciton interaction
\begin{equation} 
V_{\text{XX}}(r) = 
\begin{dcases}
 \frac{1}{4\pi\epsilon}\, \frac{e^2d^2}{r(r^2+d^2)} & \text{if } r\geq 2 R_{hc} \, ,\\
\ \ \ \ \ \infty & \text{otherwise.}
\end{dcases}
\label{Eq:V}
\end{equation} 
The hard core in $V_{\text{XX}}(r)$ of radius $R_{hc}$ takes into account the short-range two-body correlations that are strong in the low-density region where we will work \cite{Tanatar1989}. 
We determine $R_{hc}$ by comparing our results with the exciton superfluid to normal-solid transition obtained from quantum Monte Carlo (QMC) \cite{Astrakharchik2007,Boning2011}. 
The resulting value $R_{hc}\sim 0.9 r_0$ is consistent with the short-range correlation length scales determined for repulsive dipolar bosons \cite{Mora2007}.
Since the short-range two-body correlations for the supersolid should be similar to the normal solid, we use these values for $R_{hc}$ throughout.

Next we introduce variational functions for the many-particle states.
To determine the ground state, we minimize the energy for 
(i) the order parameter of the superfluid phase, 
(ii) the wave function of the exciton normal solid, and 
(iii) the order parameter of the exciton supersolid. 

(i)\ \ The order parameter of the superfluid is
\begin{equation}
\langle \hat{\Phi}_{s\!f}^\dagger(\mathbf{r})\rangle = {\Phi}_{s\!f}(\mathbf{r}) =\sqrt{\rho} \ , 
\label{Eq:SForderparameter}
 \end{equation}
normalized to $\int\! d^2 \mathbf{r}\ \lvert\sqrt{\rho}\,\rvert^2\! = \! N$.
The $\hat{\Phi}_{s\!f}^\dagger(\mathbf{r})$ creates an exciton-like boson at position $\mathbf{r}$.

(ii)\ \ For the exciton normal solid, the variational wave function is taken as a product of normalized Gaussians, each centered on a different lattice site of the exciton solid, $\mathbf{a}_i$, 
\begin{equation}
\Phi_{ns}(\mathbf{r}_1,\ldots,\mathbf{r}_N) = \prod_{i=1}^{N} 
\frac{1}{\sqrt{\pi}\sigma_{ns}}\text{e}^{-(\mathbf{r}_i-\mathbf{a}_i)^2/2\sigma_{ns}^2} \ ,
\label{Eq:ENSwavefunction}
\end{equation}
with variational parameter $\sigma_{ns}$. 
The $N$ sites $\{\mathbf{a}_i\}$, equal to the number of excitons, form a triangular lattice \cite{Wessel2005} with lattice constant $a$ determined by the exciton density, $\rho=2/(\sqrt{3}a^2)$. 
For a stable solid, $\sigma_{ns}\ll a$.

(iii)\ \ For the exciton supersolid we choose a form for the variational order parameter $\Phi_{ss}(\mathbf{r})$ \cite{Mitra2009} that 
corresponds to exactly one exciton per supersolid site, 
and contains phase coherence on the macroscopic scale,
\begin{equation}
\langle \hat{\Phi}_{ss}^\dagger(\mathbf{r})\rangle = 
\Phi_{ss}(\mathbf{r}) = \sqrt{\rho_{s\!f}}+\sqrt{\rho_{ss}}
\sum_{i=0}^{N} 
 \text{e}^{-(\mathbf{r}-\mathbf{a}_i)^2/2\sigma_{ss}^2}\ .
\label{Eq:SSorderparameter}
\end{equation}
Equation (\ref{Eq:SSorderparameter}) represents a phase coherent state as it is evident that the corresponding one-body density matrix, 
\begin{eqnarray}
\!\!\!\!\!\!\!\!\!\langle \hat{\Phi}_{ss}^\dagger(\mathbf{r})\,\hat{\Phi}_{ss}(\mathbf{r}') \rangle
 &=& N^2
(\sqrt{\rho_{s\!f}}+\sqrt{\rho_{ss}}
\sum\nolimits_i 
 \text{e}^{-(r-a_i)^2/2\sigma_{ss}^2})
 \nonumber \\
 &\times& (\sqrt{\rho_{s\!f}}+\sqrt{\rho_{ss}}
\sum\nolimits_j 
 \text{e}^{-(r'-a_j)^2/2\sigma_{ss}^2}) ,\,
 \end{eqnarray}
does not vanish when $|\mathbf{r}-\mathbf{r}'|\rightarrow \infty$, and so has off-diagonal long-range order (ODLRO) \cite{Yang1962}. 
At the same time, $\Phi_{ss}(\mathbf{r})$ and $\Phi_{ns}(\mathbf{r}_1,\ldots, \mathbf{r}_N)$ share identical diagonal long-range order from the discrete symmetry of the triangular lattice $\{\mathbf{a}_i\}$.

The normalization of the supersolid order parameter, $\int\! d^2 \mathbf{r}\ |\Phi_{ss}(\mathbf{r})|^2=N$, ensures that there is only one exciton per supersolid site. 
Thus it is not a density wave. 
There are two independent variational parameters, $\sigma_{ss}$ and $\rho_{s\!f}$.
The parameter $\rho_{ss}$ is fixed through the normalization. 

A stable supersolid requires $0<\rho_{s\!f}<\rho$.
When $\rho_{s\!f}=0$, $\Phi_{ss}(\mathbf{r})$ cannot be lower in energy than $\Phi_{ns}(\mathbf{r}_1,\ldots, \mathbf{r}_N)$, because the ODLRO in $\Phi_{ss}(\mathbf{r})$ associated with the additional overlaps of the Gaussians pushes up the exciton-exciton interaction energy \cite{Mitra2009}. 
The other limit, $\rho_{s\!f}=\rho$, gives $\rho_{ss}=0$, so $\Phi_{ss}(\mathbf{r})$ reverts to $\Phi_{s\!f}(\mathbf{r})$ for the superfluid.

We calculate the energies $\langle H \rangle$ for the three phases
with the wave functions given by Eqs.\ (\ref{Eq:SForderparameter}) -- (\ref{Eq:SSorderparameter}).
We use the effective Bohr radius, $a_B=\hbar^2\epsilon/e^2 m_e^*$, as a length scale, 
and effective mRy for energies.
In double TMD monolayers, typically $a_B\sim 0.6$ nm and mRy$\sim 0.2$ meV. 

\begin{figure}[t]
\centering
\includegraphics[trim={0.23cm 0.5cm 0.48cm 0.72cm},clip=true,width=\columnwidth]{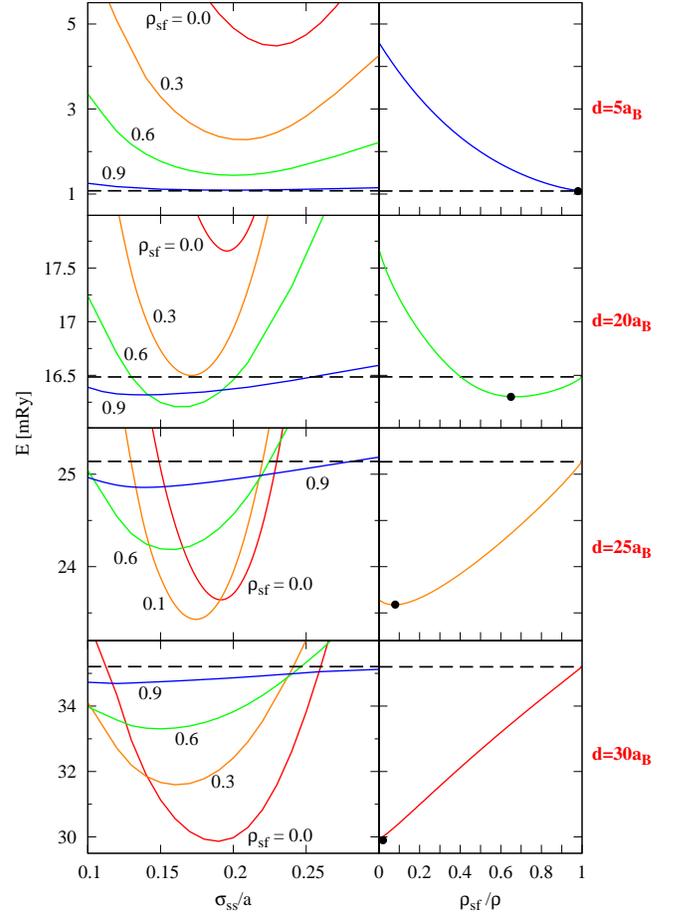}
\caption{
Exciton supersolid energy $E_{ss}$ for layer separations $d$ as indicated, for a fixed density at $r_0=30 a_B$.
Dashed line shows for reference the superfluid energy $E_{s\!f}$.
Left panels: $E_{ss}$ as a function of the localization parameter $\sigma_{ss}/a$, for fixed values of the superfluid component $\rho_{s\!f}/\rho$ as labelled. 
Right panels: minimum in $\sigma_{ss}$ of $E_{ss}$ (cf.\ left panels), plotted as a function of the superfluid component $\rho_{s\!f}/\rho$.
The dots highlight the absolute minimum of $E_{ss}$.
} 
\label{Fig:ESS_rhosf}
\end{figure}
\begin{figure*}[ht]
\centering
\includegraphics[width=0.95\textwidth]{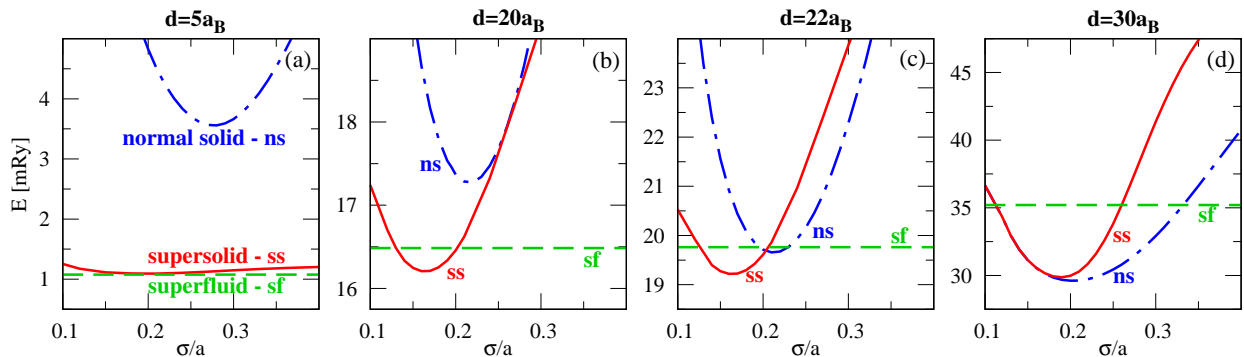}
\caption{
Energies of the superfluid ($s\!f$), exciton supersolid ($ss$) and exciton normal solid ($ns$) phases, as labelled, for different layer separations $d$. 
Density fixed at $r_0=30 a_B$.
The dependence of $E_{ss}$ and $E_{ns}$ on the localization variational parameter, $\sigma=\sigma_{ss}$ or $\sigma=\sigma_{ns}$, is shown. 
Each $E_{ss}$ curve uses the value of $\rho_{s\!f}/\rho$ yielding the lowest energy.}
\label{Fig:Energies_beta}
\end{figure*}

Figure \ref{Fig:ESS_rhosf} compares the supersolid energy $E_{ss}$ (solid lines) with the superfluid energy $E_{s\!f}$ (dashed lines), for layer separations $d$ at a fixed density corresponding to $r_0=30 a_B$. 
$E_{ss}$ is shown as a function of the two variational parameters for the supersolid, $\rho_{s\!f}/\rho$ and $\sigma_{ss}/a$.
We recall that, by definition, $E_{ss}=E_{s\!f}$ when $\rho_{s\!f}/\rho=1$.

For the smallest layer separation shown, $d=5 a_B$, the minimum in $E_{ss}$ as a function of $\sigma_{ss}$ increases monotonically as $\rho_{s\!f}/\rho$ decreases from $1$ to $0$. 
The monotonic increase implies that the superfluid energy forms a lower bound on the exciton supersolid energy.
So for $d=5 a_B$, the supersolid is not the ground state. 
In contrast, for a larger $d=20 a_B$, the minimum in $E_{ss}$ as a function of $\sigma_{ss}$ first decreases as $\rho_{s\!f}$ decreases from $1$. 
It drops below $E_{s\!f}$, reaching a lowest value at $\rho_{s\!f}/\rho=0.6$. 
Thus, the exciton supersolid is more stable than the superfluid for $d=20 a_B$.
As $d$ is further increased, this minimum in $E_{ss}$ reaches a lowest value at smaller and smaller values of $\rho_{s\!f}/\rho$.
By $d=25 a_B$, the minimum is at $\rho_{s\!f}/\rho=0.1$, and by $d=30 a_B$ it decreases monotonically all the way down to $\rho_{s\!f}/\rho=0$.
Again, for these cases the supersolid is more stable than the superfluid.

Figure \ref{Fig:Energies_beta} shows for the same density as Fig.\ \ref{Fig:ESS_rhosf}, the energies of the exciton superfluid $E_{s\!f}$, supersolid $E_{ss}$, and normal solid $E_{ns}$ phases. 
The dependence of $E_{ns}$ on $\sigma=\sigma_{ns}$ and $E_{ss}$ on $\sigma=\sigma_{ss}$ is plotted. 
Each $E_{ss}$ curve uses the value of $\rho_{s\!f}/\rho$ yielding the lowest energy (Fig.\ \ref{Fig:ESS_rhosf}).
For $d=5 a_B$, the minimum of $E_{ns}$ lies above $E_{s\!f}$,
so the superfluid is the ground state. 
By $d=20 a_B$, the minimum of $E_{ss}$ has dropped below $E_{s\!f}$ and it is still below the minimum of $E_{ns}$, so the supersolid is the ground state. 
$E_{ns}$ finally touches $E_{s\!f}$ at $d\simeq 22 a_B$ (Fig.\ \ref{Fig:Energies_beta}(c)), matching the position in phase space of the exciton liquid to normal-solid transition predicted by QMC in Ref.\ [\citenum{Astrakharchik2007}]. 
However since for this $d$ the supersolid phase is still the ground state, it preempts this transition. 
Only when the layer separation is increased to $d=30 a_B$, does the minimum of $E_{ns}$  drop below $E_{ss}$. 
This signals an intriguing supersolid to normal-solid transition, in which the ground-state crystal structure is unchanged but the quantum phase coherence collapses. 
We note that the energies around this density and layer separation are consistent with QMC calculations \cite{DePalo2002}. 

It is interesting to compare in Fig.\ \ref{Fig:Energies_beta}, the dependence of $\sigma/a$ at the minima of $E_{ns}$ and $E_{ss}$ on the layer spacing. 
As $d$ increases, the repulsive interaction (Eq.\ (\ref{Eq:V})) becomes stronger than the kinetic energy, and the particles should become more localized on their lattice sites, corresponding to a decreasing $\sigma/a$. 
This is clearly visible in the minimum of the normal solid. 
In contrast for the supersolid, the localization degree $\sigma/a$ remains notably little changed over this range of $d$.

\begin{figure}[t]
\centering
\includegraphics[trim={0.6cm 1.5cm 0.05cm 0},clip=true,width=\columnwidth]{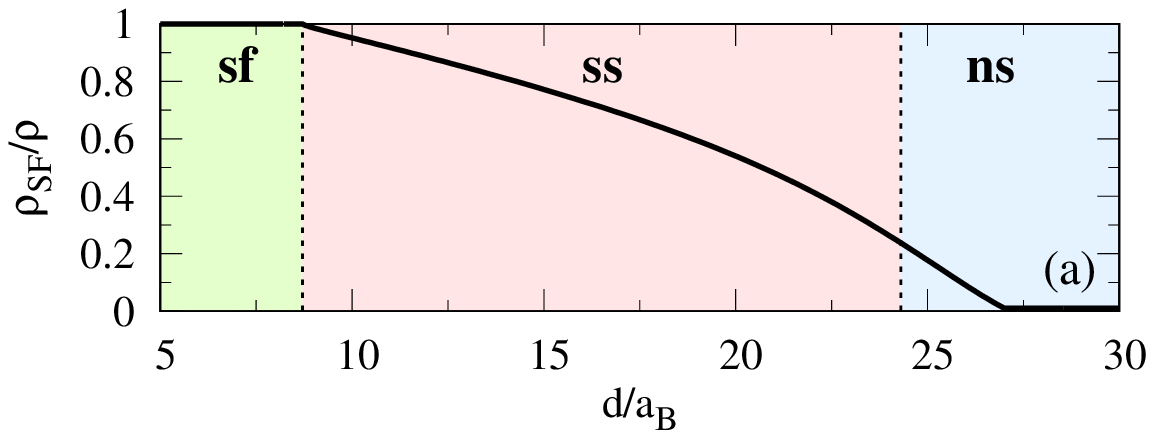}\\
\vspace{-0.465cm}
\includegraphics[trim={0cm 0.5cm 0cm 0},clip=true,width=\columnwidth]{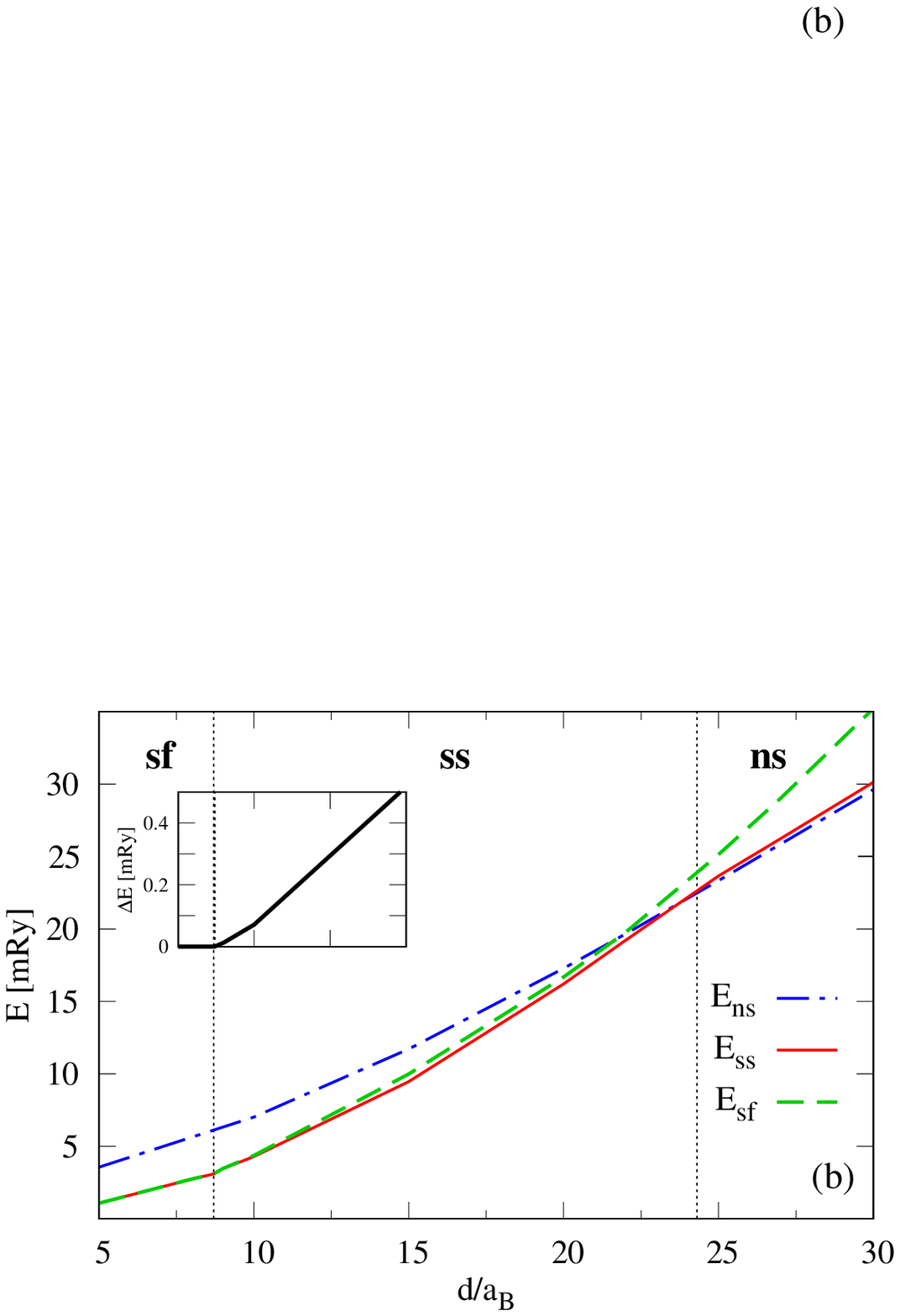}
\caption{
a) Dependence of $\rho_{s\!f}/\rho$ on $d$ for the lowest value of $E_{ss}$. 
Coloured areas indicate the ground state: $s\!f$ - superfluid, $ss$ - exciton supersolid, $ns$ - exciton normal solid. 
b) Minimum superfluid energy $E_{s\!f}$, exciton supersolid energy $E_{ss}$, and exciton normal solid energy $E_{ns}$ as functions of $d$. 
The crossings of energies are marked by vertical dotted lines.
Inset shows energy difference $\Delta E = E_{s\!f} - E_{ss}$ on same $d$ scale.
Fixed density, with $r_0=30 a_B$.
}
\label{Fig:rhosf_d}
\end{figure}
Figure \ref{Fig:rhosf_d}(a) shows that for the supersolid it is $\rho_{s\!f}/\rho$ that changes dramatically with $d$. 
The figure plots $\rho_{s\!f}/\rho$ for the lowest $E_{ss}$ as a function of $d$, for the same density as Fig.\ \ref{Fig:ESS_rhosf}. 
The ground states (shaded areas) are extracted from Fig.\ \ref{Fig:rhosf_d}(b), which shows the minima of the energies of the three phases.
The two energy crossings, corresponding to the superfluid--supersolid and supersolid--normal solid transitions, are indicated by the dotted lines.
For small $d$, $\rho_{s\!f}/\rho=1$ and the superfluid is the ground state.
For intermediate values of $d$, $\rho_{s\!f}/\rho$ decreases, but before it reaches zero, there is the transition to the normal solid ground state, with its minimum energy dropping below the minimum supersolid energy. 
We find that for all densities, the supersolid phase is only stable relative to the normal solid when $\rho_{s\!f}/\rho\gtrsim 0.25$, confirming that because of the ODLRO, a stable supersolid only exists for non-zero $\rho_{s\!f}/\rho$. 
In other words, a superfluid component in the supersolid order parameter is necessary for stability.

\begin{figure}[t]
\centering
\includegraphics[trim={0.5cm 0.4cm 0.45cm 0.1cm},clip=true,width=\columnwidth]{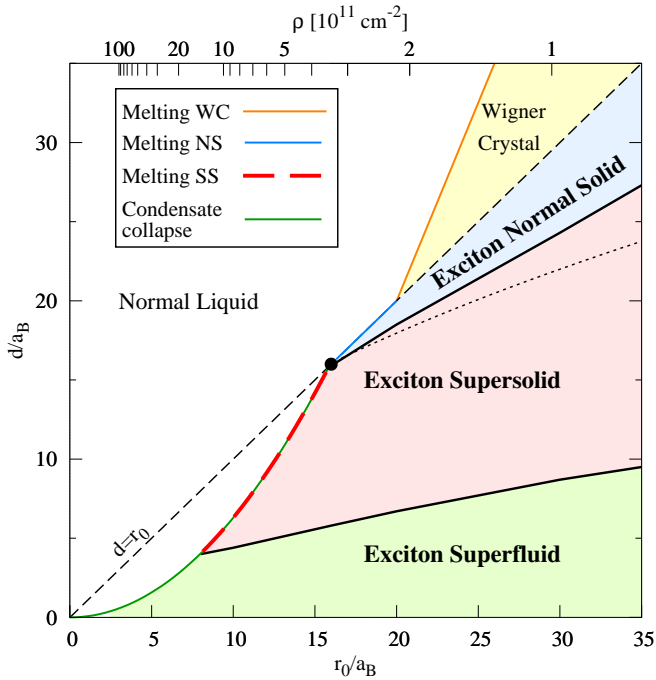}
\caption{Phase diagram at zero temperature. 
$d$ is the layer separation and $r_0$ characterizes the density $\rho$.
Dotted line is the transition to the exciton normal solid predicted in Ref.\ \cite{Astrakharchik2007}. 
}
\label{Fig:Phasediagram}
\end{figure}

Figure \ref{Fig:Phasediagram} shows the phase diagram.
The energies driving the phases are 
the electron-hole attraction between layers $V_{eh}=-e^2/(4 \pi \epsilon \,d)$, 
the repulsion between charges within each layer $V_{ee}=V_{hh}=e^2/(4 \pi \epsilon \, r)$, 
the effective exciton-exciton repulsion $V_{\text{XX}}$ (Eq.\ (\ref{Eq:V})),
and the Fermi energy $E_F= \hbar^2/(2 m_e^* r_0^2)$. 

The dashed line $d=r_0$ divides the phase diagram 
into a lower region where the average interlayer attraction $\langle V_{eh}\rangle$ is the most important,
and an upper region where the average intralayer repulsions $\langle V_{ee}\rangle = \langle V_{hh}\rangle $ are the most important.
For $d\ll r_0$, due to the electron-hole attraction $V_{eh}$, the exciton-exciton repulsion is dipolar, $V_{\text{XX}}(r) = e^2d^2/(4\pi\epsilon\, r^3)$. 
For $d\gg r_0$, the effect of $V_{eh}$ is negligible and the exciton-exciton repulsion becomes Coulombic, $V_{\text{XX}}(r) = 2e^2/(4\pi\epsilon\, r)$.

The phase space region for small separations $d$ and large $r_0$ where the ground state is superfluid (green area), has been intensively studied theoretically and experimentally \cite{Perali2013,LopezRios2018,Burg2018,Wang2019}.
If $r_0$ is decreased, $V_{eh}/E_F\sim r_0^2/(d a_B)$ will decrease, until the superfluid gap generated by $V_{eh}$ drops below $E_F$. 
At this point, the strong screening suppresses the superfluid gap so that at $V_{eh}/E_F\simeq 17$ the condensate collapses and there is a transition to the normal-state liquid \cite{LopezRios2018} (green line). 

At larger $d$, there is the new supersolid phase occupying a large region of phase space (red area). 
The supersolid transition occurs much earlier than the liquid to normal-solid transition predicted at $V_{\text{XX}}/E_F = d^2/(r_0 a_B)\simeq 17$ \cite{Astrakharchik2007,Boning2011} (dotted line). 
The appearance of the supersolid well below the $d$ predicted for the normal solid is due to the presence of $\rho_{s\!f}>0$ in the supersolid order parameter that stabilizes the supersolid phase. 
This phase persists for $V_{\text{XX}}/E_F > 17$, until there is a transition from supersolid to the normal solid (blue area). 
Further increasing $d$, we cross $d=r_0$ where $V_{\text{XX}}$ becomes Coulombic. 
Here a low-density bilayer Wigner Crystal ground state has been predicted \cite{Szymanski1994} (yellow area).

We now turn to the melting of the solid phases at high densities. 
The melting of the bilayer Wigner Crystal was determined in Ref.\ [\citenum{DePalo2002}] (orange line). 
As for the exciton solids, above $d=r_0$ a Coulombic-like $V_{\text{XX}}(r)$ cannot support a solid at such high densities. 
In this way, the exciton normal-solid melting line is approximately at $d=r_0$ \cite{Boning2011} (blue line).

For the supersolid phase there is a second melting mechanism. 
If the supersolid quantum phase coherence collapses at a $d$ value for which $V_{\text{XX}}/E_F < 17$, then the exciton repulsion will not be strong enough to support a normal solid, and so it will melt. 
The mechanism for this phase coherence collapse is expected to be the same as for the superfluid, that is, a strong increase in the screening. 
Therefore, we extend the calculation of the superfluid condensate collapse to approximately determine the collapse of the supersolid quantum phase coherence, and so also the supersolid melting (red line). 
The approximation is justified both by the resulting supersolid melting line lying close to $d=r_0$, and also because we find the superfluid component $\rho_{s\!f}/\rho$ in the supersolid is large there.

We note there is a triple point at the intersection of the supersolid melting, the supersolid to normal-solid transition, and the normal-solid melting. 
The ever present disorder in an experiment will spread the triple point over an area of phase space with co-existing supersolid, normal solid and normal liquid domains, with exciting physics stemming from the diverse and exotic interfaces separating these domains.
A fascinating possibility would be Josephson tunneling between supersolid puddles embedded in a normal-state background. 

In conclusion, the supersolid ground state of excitons that we are predicting is robust and extends over a wide area of the equilibrium phase diagram.
Since there is precisely one exciton occupying each supersolid site, our supersolid is of the Chester type and fundamentally differs from proposals of supersolids that are superfluids with a periodic density modulation, resembling density waves.
Intriguingly, we find that a superfluid component in the supersolid order parameter is essential to stabilize the supersolid, at least in the absence of other stabilizing factors such as vacancies \cite{Kurbakov2010}.
By changing the length of the exciton dipole moment (layer separation) relative to the exciton spacing (density), the exciton-exciton interactions can be tuned to stabilise the superfluid, supersolid, or normal solid. 
The necessary ranges of densities, layer separations, and dielectric constants are readily accessible experimentally and controllable in semiconductor heterostructures. 
In addition, our augmented phase diagram offers a rich selection of novel phenomena in the vicinity of the triple point, the solid-solid transition associated with the loss of quantum phase coherence, the supersolid melting coinciding with the condensate collapse, etc., all worthy of further investigation.

We thank Davide Galli, Carlos S\'a de Melo and Jacques Temp\`ere for useful discussions.
The work was supported by the Flemish Science Foundation (FWO-Vl), and by the Australian Government through the Australian Research Council Centre of Excellence in Future Low-Energy Electronics (Project No. CE170100039).

\end{document}